\documentclass[preprint]{aastex63}
\usepackage{color}
\usepackage{amsmath}

\begin{document}

\title{The rarity of repeating fast radio bursts from binary neutron star mergers}

\author[0000-0001-6545-4802]{G. Q. Zhang}
\affiliation{School of Astronomy and Space Science, Nanjing University, Nanjing 210093, China}

\author{S. X. Yi}
\affiliation{School of Physics and Physical Engineering, Qufu Normal University, Qufu 273165, China}

\author[0000-0003-4157-7714]{F. Y. Wang}
\affiliation{School of Astronomy and Space Science, Nanjing University, Nanjing 210093, China}
\affiliation{Key Laboratory of Modern Astronomy and Astrophysics (Nanjing University), Ministry of Education, Nanjing 210093, China}

\correspondingauthor{F. Y. Wang}
\email{fayinwang@nju.edu.cn}

\begin{abstract}
Fast radio bursts (FRBs) are extragalactic, bright pulses of
emission at radio frequency with milliseconds duration.
Observationally, FRBs can be divided into two classes, repeating
FRBs and non-repeating FRBs. At present, twenty repeating FRBs have
been discovered with unknown physical origins. Localization of the
first repeating FRB 121102 and discovery of an associated persistent
radio source support that FRBs are powered by young millisecond
magnetars, which could be formed by core-collapses of massive stars
or binary neutron stars mergers. These two formation channels can be
distinguished by gravitational waves generated by binary neutron
stars mergers. We first calculate the lower limit of the local
formation rate of repeating FRBs observed by the Canadian Hydrogen
Intensity Mapping Experiment (CHIME) . Then we show that only a
small fraction ($6\%$) of repeating FRBs is produced by young
magnetars from binary neutron star mergers, basing on the
gravitational wave detections by the third observing run (O3) of
Advanced LIGO/Virgo gravitational-wave detectors. Therefore, we
believe that repeating FRBs are more likely produced by the
magnetars newborn from the core-collapses of massive stars rather
than the magnetars from the binary neutron stars mergers.
\end{abstract}

\section{Introduction \label{sec:intro}}
Fast radio bursts (FRBs) are millisecond duration radio pulses with
large dispersion measures (DMs) exceed the Milky Way contribution
along the line of sight \citep{Cordes19,Petroff19,Platts19}. At
present twenty FRBs are found to show multiple bursts
\citep{Spitler16,CHIME2019Natur.566..235C,Kumar19,CHIME2019,Fonseca2020arXiv200103595F},
indicating non-cataclysmic sources. Whether or not all FRBs are
repeating sources is still under debate
\citep{Caleb2019MNRAS.484.5500C,James19}. The rate of non-repeating
FRBs is larger than that of any known cataclysmic events, pointing
to a population of repeating sources
\citep{Ravi2019NatAs.tmp..405R}. The physical origins of FRBs are
still mysterious. The observed short timescale structure in FRB
light curves requires stellar-mass compact objects as central
engines, such as magnetars
\citep{Popov2013arXiv1307.4924P,Kulkarni14,Murase16,
Metzger2017ApJ...841...14M,Beloborodov17,Margalit2018MNRAS.481.2407M},
binary neutron stars mergers
\citep{Totani13,Wang16,Yamasaki18,Zhang20}, and interacting models
\citep{Geng15,Dai16,Zhang2017ApJ...836L..32Z}.

The first repeating FRB source, FRB 121102, has been localized to a
star-forming, dwarf galaxy at $z=0.19$
\citep{Chatterjee2017Natur.541...58C,Tendulkar2017ApJ...834L...7T}.
This FRB is also spatially associated with a luminous persistent
radio source \citep{Marcote2017ApJ...834L...8M}. The properties of
the host galaxy of FRB 121102 are similar to those of superluminous
supernovae and long gamma-ray bursts
\citep{Tendulkar2017ApJ...834L...7T,Metzger2017ApJ...841...14M,Zhang2019MNRAS.487.3672Z},
which supports the central engine of repeating FRBs being a young
magnetar formed by core-collapse of a massive star
\citep{Metzger2017ApJ...841...14M,Margalit2018MNRAS.481.2407M}. In
this scenario, the persistent radio source could be produced by
emission from a compact magnetized nebula surrounding the young
magnetar
\citep{Metzger2017ApJ...841...14M,Kashiyama2017ApJ...839L...3K,Margalit2018MNRAS.481.2407M}.
Recently, three FRBs (FRBs 180924, 181112 and 190523) that have not
yet been observed to repeat are localized
\citep{Bannister2019Sci...365..565B,Ravi2019Natur.572..352R,Prochaska2019Sci...366..231P}.
The host galaxies of these three FRBs are massive galaxies with
little star formation, which are similar to those of short gamma-ray
bursts \citep{Berger2014ARA&A..52...43B}. Moreover, the offsets
between the bursts and host centers are about 4 kpc and 29 kpc for
FRB 180924 and FRB 190523, respectively. Therefore, the central
magnetar powering FRBs may also be formed by binary neutron star
(BNS) mergers \citep[Wang et al. 2019]{Margalit19}. The properties
of some repeating FRBs are consistent with the scenario of a
millisecond magnetar formed by a BNS merger
\citep{Margalit19,Wang19,CHIME2019}. From numerical simulation,
\citet{Yamasaki18} found that a fraction of BNS mergers may leave a
rapidly rotating and stable neutron star, and it may generate
repeating FRBs in 1-10 years after the merger.

The two formation channels of magnetars, i.e., core collapse or BNS
merger, can be distinguished by the rate of gravitational-wave (GW)
events observed \citep{Zhang14,Callister16} by Advanced LIGO/Virgo.
In this paper, we constrain the fraction of repeating FRBs from BNS
mergers with gravitational wave observations. In Section 2 we
describe a robust method to calculate the lower limit of the local
rate of repeating FRBs. The results are shown in Section 3.
Discussion and conclusions are given in Section 4.

\section{Method}
We collect the data of repeating FRBs from the observation of CHIME
\citep{CHIME2019Natur.566..235C,CHIME2019,Fonseca2020arXiv200103595F},
which is the largest sample of repeating FRBs observed by the same
telescope. \citet{CHIME2019Natur.566..235C} reported the second
repeating FRB 180814 discovered during the commissioning phase and
\citet{CHIME2019} also reported eight new repeating sources observed
from August 28, 2018 to March 13, 2019. Recently, nine new repeating
FRBs were reported by \citet{Fonseca2020arXiv200103595F}. These new
FRBs are observed from August 28, 2018 to September 30, 2019. The
total DM of each FRB and DM contributed by the Milky Way have been
given in their works. We use these data to estimate the redshifts of
these repeating FRBs. In the following, we adopt the method
developed by \citet{Ravi2019NatAs.tmp..405R} to investigate the
lower limit of the formation rate.

The DMs of FRBs can be separated into different parts
\begin{equation}
\label{eq:DM}
\rm DM = DM_{\mathrm{MW}} + DM_{\mathrm{halo}} + DM_{\mathrm{IGM}} + \frac{DM_{\mathrm{host}}}{1 + z},
\end{equation}
where DM$_{\mathrm{MW}}$, DM$_{\mathrm{halo}}$, DM$_{\mathrm{IGM}}$
and DM$_{\mathrm{host}}$ represent the DM contributed by the Milky
Way, the Milky Way halo, the intergalactic medium (IGM), and the
host galaxy, respectively. In our analysis, the values of
DM$_{\mathrm{MW}}$ are derived from the NE2001 model
\citep{Cordes2002astro.ph..7156C}. The DM$_{\mathrm{halo}}$ is
difficult to estimate. A uniform distribution of
DM$_{\mathrm{halo}}$ from 50 pc cm$^{-3}$ to 80 pc cm$^{-3}$ is
adopted in our work \citep{Prochaska2019MNRAS.485..648P}. In order
to have a good description of DM$_{\mathrm{host}}$, we use the
results of Zhang, Yu \& Wang (2019). They use the IllustrisTNG
simulation to estimate the value of DM$_{\mathrm{host}}$ at
different redshifts. DM$_{\mathrm{IGM}}$ can be calculated from
\citep{Ioka2003ApJ...598L..79I,Deng2014ApJ...783L..35D}
\begin{equation}
\label{eq:DMigm}
\mathrm{DM}_{\mathrm{IGM}}(z)=\frac{3 c H_{0}
    \Omega_{\mathrm{b}}}{8 \pi G m_{\mathrm{p}}} f_{\mathrm{IGM}}
\int_{0}^{z} \frac{H_{0} f_{\mathrm{e}}\left(z^{\prime}\right)
    \left(1+z^{\prime}\right)}{H\left(z^{\prime}\right)} \mathrm{d} z^{\prime},
\end{equation}
where $ H_0$ is the Hubble constant, $ \Omega_{\mathrm{b}} $ is
baryon density, $ m_p $ is the rest mass of protons, $ H(z) $ is the
Hubble parameter, $f_{\mathrm{IGM}} \simeq 0.83$ is the fraction of
baryon mass in the IGM \citep{Shull12} and
\begin{equation}
    \label{eq:fe}
    f_e(z')  = y_1 f_{e,H}(z') + y_2 f_{e,He}(z')
\end{equation}
is the number ratio between the free electrons and baryons. In this
equation, $ f_{e,H}(z') $ and $ f_{e,He}(z') $ are the ionization
fractions of hydrogen and helium, respectively. $ y_1 \simeq 3/4$ is
the fraction of hydrogen in the universe and $ y_2 \simeq 1/4 $ is
the helium abundance. Assuming the hydrogen and helium are
fully-ionized, we can get $ f_e(z') \simeq 7/8 $.

\begin{figure}
    \centering
    \includegraphics[width=0.7\linewidth]{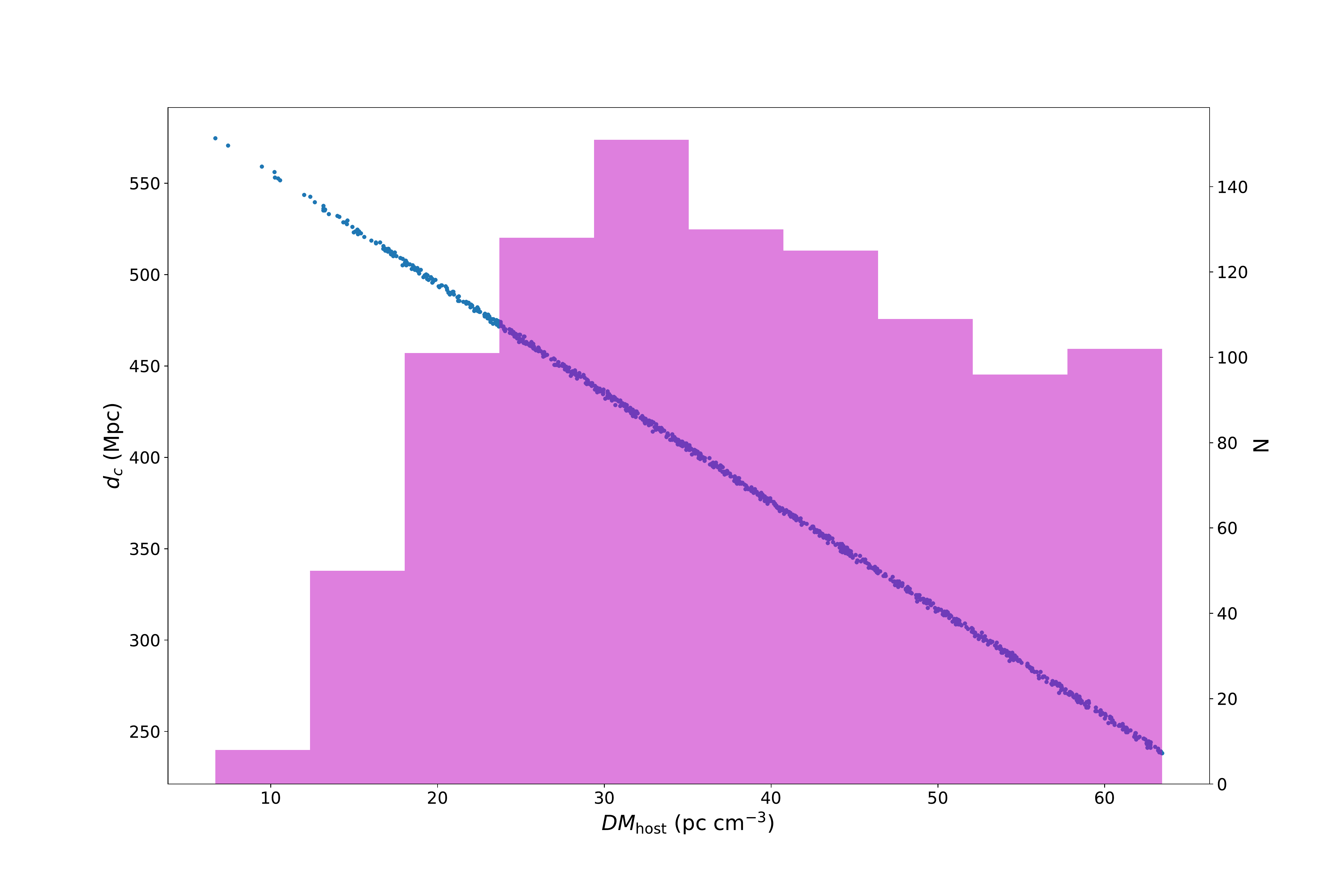}
    \caption{The red histogram shows 1000 host galaxy dispersion measures
        randomly selected from $10^5$ simulated host galaxy dispersion measures according to Zhang, Yu \& Wang (2019).
        The number of host galaxies is shown in the right axis. The value of DM$_{\mathrm{host}} $ satisfies
        DM$_{\mathrm{host}} < \rm DM_{\mathrm{ex,lowest}} $, where DM$_{\mathrm{ex,lowest}} $
        is the lowest DM$_{\mathrm{ex}} $ among repeating FRBs. In this case, only three FRBs with lowest DM$_{\mathrm{ex}} $ are considered.
        The blue scatter points represent critical distances $d_c$ for different values of host galaxy dispersion measures. The critical distance $d_c$
        denotes the distance to include three FRBs with the lowest DM$_{\mathrm{ex}} $.}
    \label{fig:dmdis}
\end{figure}

The observation by CHIME is incomplete for FRBs with high values of
dispersion measures
\citep{Shull2018ApJ...852L..11S,Ravi2019NatAs.tmp..405R}. In order
to overcome this incompleteness, the local formation rate of
repeating FRBs is derived from the lowest-DM CHIME FRBs, which is
similar to that of \citet{Ravi2019NatAs.tmp..405R}. Three FRBs with
lowest extragalactic dispersion measures DM$_{\mathrm{ex}}$ are
chosen for analysis, where $\rm DM_{\mathrm{ex}} =
DM_{\mathrm{halo}} + DM_{\mathrm{IGM}} + DM_{\mathrm{host}} / (1 +
z)$. Since the chosen FRBs have very low redshifts, the effects of
redshift evolution are negligible. We introduce a probability
function $p(<d|DM_{\rm ex})$ to indicate the possibility that an FRB
with DM$_{\rm ex}$ has the distance $d_{\rm FRB} < d$. The critical
distance $d_c$ is determined by the following criteria: the two FRBs
with lowest DM$_{\mathrm{ex}}$ have $p(<d_c|DM_{\mathrm{ex}}) \simeq
1$ and the FRB with third-lowest DM$_{\mathrm{ex}}$ has
$p(<d_c|DM_{\mathrm{ex}}) \simeq 0.95$. The probability
$p(<d|DM_{ex})$ can be derived from the distributions of
DM$_{\mathrm{host}}$, DM$_{\mathrm{halo}}$ and DM$_{\mathrm{IGM}}$.
In our analysis, we assume that the distribution of
DM$_{\mathrm{halo}}$ satisfies the uniform distribution between 50
pc cm$^{-3}$ and 80 pc cm$^{-3}$
\citep{Prochaska2019MNRAS.485..648P}. The distribution of
DM$_{\mathrm{IGM}}$ satisfies the standard normal distribution with
1$\sigma$ error of 10 pc cm$^{-3}$ \citep{Shull2018ApJ...852L..11S}.
According to the results of Zhang, Yu \& Wang (2019), 
we simulate $10^5$ host galaxy dispersion measures and randomly
select 1000 host galaxy dispersion measures which satisfy
DM$_{\mathrm{host}} < \mathrm{DM}_{\mathrm{ex, lowest}} $. The
histogram of DM$_{\mathrm{host}}$ is shown in Figure
\ref{fig:dmdis}. The red histogram shows the distribution of
DM$_{\mathrm{host}}$. Based on the assumptions of
DM$_{\mathrm{IGM}}$ and DM$_{\mathrm{halo}}$, we derive the critical
distance $d_c$ for different values of DM$_{\mathrm{host}}$ and show
the results as blue points in Figure \ref{fig:dmdis}. The blue
points denote the value of $ d_c $ for different host galaxy
dispersion measures. If there is no DM$_\mathrm{host}$ contribution,
the value is $d_c \simeq 574 $ Mpc. By comparison, the value of
$d_c$ is about 739 Mpc for non-repeating FRBs of CHIME without
DM$_\mathrm{host}$ contribution \citep{Ravi2019NatAs.tmp..405R}.

\section{Results}
\begin{figure}
    \centering
    \includegraphics[width=0.7\linewidth]{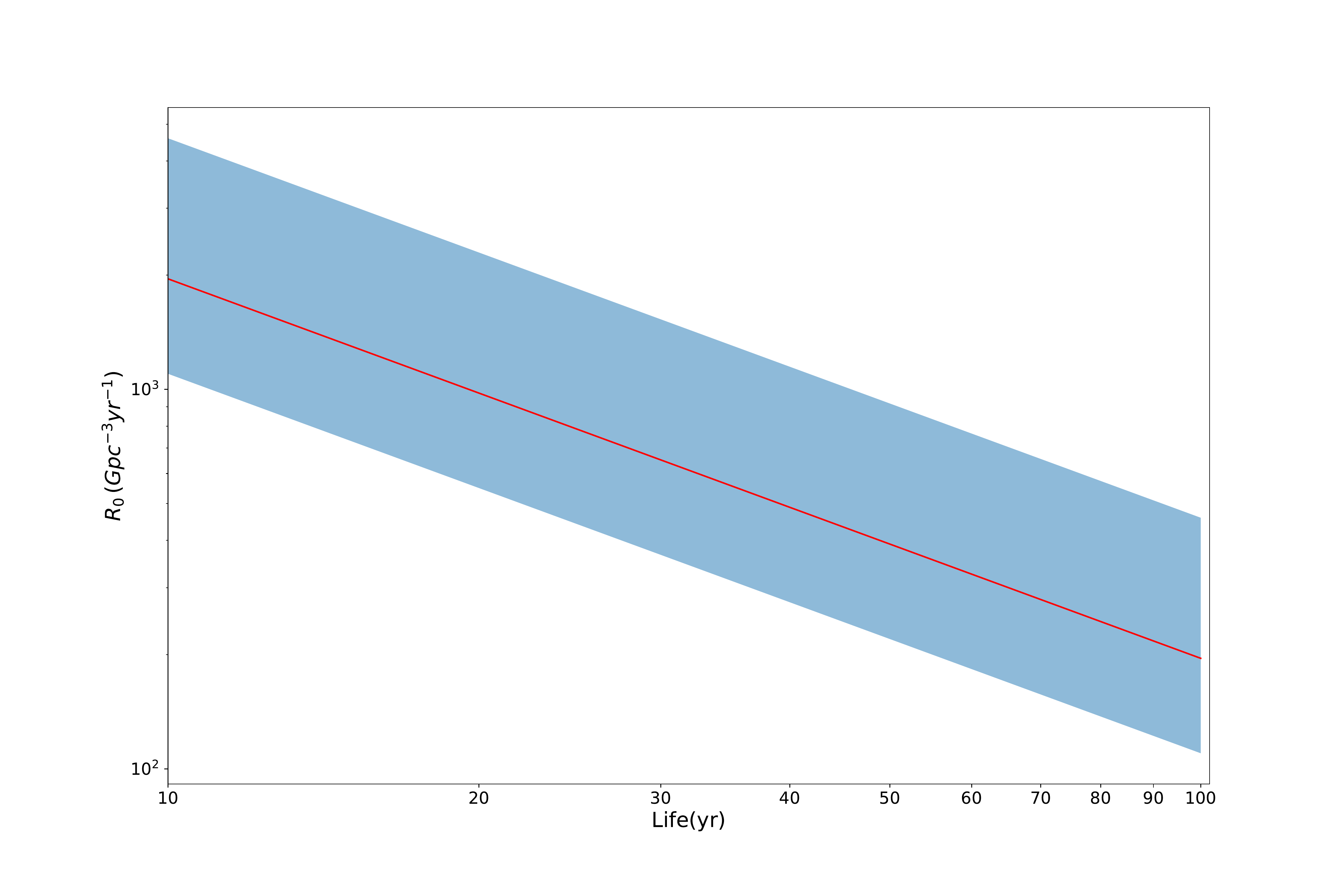}
    \caption{The red line indicates the lower limit of
        volumetric rate $R_0$ given a $d_c$ corresponding to three CHIME repeating FRBs with the lowest-DM$_\mathrm{ex}$. The blue area represents
        1-$\sigma$ confidence level caused by the uncertainties of DM$_{\mathrm{host}}$.
        The beaming factor $f_b=0.1$ is used. The lower limit of formation rate is higher than the
        rate of GW events detected by the aLIGO/Virgo.}
    \label{fig:rate}
\end{figure}

For the three FRBs with lowest DM$_{\rm ex}$, we calculate the lower
limit of formation rate $R_0$ from
\begin{equation}
\label{eq:rate}
R_0 \tau = \frac{N}{V(d_c)\eta_\Omega f_b},
\end{equation}
where $\tau$ is the lifetime of the repeater, $ N = 3 $ is the
number of FRBs, $V(d_c) = 4\pi d_c^3 / 3$ is the largest volume that
contains the three FRBs, $\eta_\Omega$ is the sky coverage of CHIME,
and $f_b$ is the beaming factor. In this equation, we introduce the
lifetime of repeaters $ \tau $. If the observation time $T_{obs}\ll
\tau$, the number of repeating FRBs observed in $T_{obs}$ is
actually the sum of repeating FRBs that burst in the past $ \tau $
years. In Equation (\ref{eq:rate}), the left part is the predicted
number density of FRBs and right part is the observed number density
of repeating FRBs.
\cite{CHIME2019} reported eight new repeating FRBs in the
observation during the intervals from August 28, 2018 to March 13,
2019. The latest nine repeating FRBs were observed from August 28,
2018 to September 30, 2019. The observation time is about 400 days.
This observation time is so long that we can ignore the effect of
the active periods of the repeating FRBs. Besides, compared with the
lifetime of repeaters, the observation time is short. Therefore, we
can use Equation (\ref{eq:rate}) to derive the lower limit of the
formation rate.
The field of view of CHIME is about $256$ deg$^2$
\citep{CHIME2018ApJ...863...48C}, so $\eta_\Omega = 256 / 41242.96$.
Based on the above arguments, the lower limit of the local formation
rate is
\begin{equation}
\label{eq:ratere} R_0 = 1955 ^{+2633}_{-856} f_b^{-1} \tau^{-1}
\mathrm{Gpc}^{-3} \mathrm{yr}^{-1}.
\end{equation}
The 1$\sigma$ error comes from the
uncertainty of DM$_{\mathrm{host}}$.
The lifetime of repeater significantly affects the value of $R_0$.
The typical magnetic active timescale is 20 years for high-mass
neutron stars and 700 years for normal-mass neutron stars
\citep{Beloborodov16}. For Galactic BNS systems, the mass
distribution peaks above the maximum stable mass
\citep{Margalit19a}. If extragalactic population has similar mass
distribution, a large fraction of mergers will leave high-mass
neutron stars. Furthermore, a merger-remnant magnetar may emit
repeating bursts for about 10 years from numerical simulation
\citep{Yamasaki18}, which is consistent with the active time of FRB
121102. Therefore, the most probable lifetime for magnetic activity
is 20 years. From the FRB 121102 observation, the constraint on the
age of centeral magnetar is also about a few decades
\citep{Kashiyama2017ApJ...839L...3K,Metzger2017ApJ...841...14M,Cao17}.
Therefore, we adopt the lifetime between 10 and 100 years. The value
of $f_b$ is not constrained observationally. We adopt a fiducial
value of $f_b=0.1$ \citep{Nicholl2017ApJ...843...84N}, which is
appropriate for pulsars. For illustration purpose, we calculate the
local formation rate $R_0$ for lifetime from 10 to 100 years. The
result is shown as red line in Figure \ref{fig:rate}. The blue
region is 1$ \sigma $ confidence level.
In this framework, the formation rate of repeating FRBs
is

\begin{equation}
R_0 \sim
\begin{cases}
1955 ~\rm Gpc ^{-3}  yr ^{-1}, \, \tau=10 \, yr
\\
195  ~ \rm Gpc ^{-3}  yr ^{-1}, \, \tau=100\, yr
\end{cases}
.
\label{R0}
\end{equation}

BNS mergers will produce gravitational waves, which can be observed
by the Advanced LIGO/Virgo gravitational-wave detectors. Therefore,
the fraction of magnetars born in BNS mergers can be constrained by
gravitational wave observations. Until 1 February 2020, the O3
observation of Advanced LIGO/Virgo has been carried out for about
300 days. Only one event S190425Z has a probability of more than
99\% produced by BNS
merger\footnote{\url{https://gracedb.ligo.org/latest/}}.
If the lifetime of repeating sources is 10 years, considering the
observation time $T = 300$ days and the detection range is 170 Mpc,
the Advanced LIGO/Virgo may detect $33.08^{+5.75}_{-5.75}$
gravitational-wave events from BNS mergers, where the error is
Poisson error.
If the lifetime is 100 years, We expect the Advanced LIGO/Virgo
detect $3.31^{+1.82}_{-1.82}$ GW events. Considering the most
probable lifetime for magnetic activity is $ \tau \simeq 20 $ years,
the Advanced LIGO/Virgo can detecte $ 16.54 \pm 4.07 $ GW events in
300 days. If the detection of GW events satisfies the Poisson
distribution, the probability that the number of observed
gravitational-wave events less than three is about $ 6 \times
10^{-5} $, which is almost impossible. Considering one GW event from
BNS merger was observed in the O3 observation, the fraction of
repeating FRBs from BNS mergers is only 6\%.

\section{Discussion and Conclusions}

In above calculations, the NE2001 model of dispersion measures for
Milky Way is used. Below, we consider the YMW17 model for Milky Way
\citep{Yao17}. The values of DM$_{\rm ex}$ are comparable in the two
models, except for FRB 180916.J0158+65. In this case, the value of
DM$_{\rm ex}$ for FRB 180916.J0158+65 is only about 20 pc cm$^{-3}$,
which is the lowest one. Compared with DM$_{\rm ex}$=149 pc
cm$^{-3}$ for this FRB in the NE2001 model, the corresponding
critical distance $d_c$ is much smaller than that in the NE2001
model. The above volumetric formation rate of repeating FRBs is
conservative because some CHIME FRBs that have not been observed to
repeat may be intrinsically repeating
\citep{Ravi2019NatAs.tmp..405R}. Therefore, the derived lower limit
is conservative and robust.

It is well known that BNS merger is the leading model for short
gamma-ray bursts, as confirmed by the first BNS gravitational-wave
event GW170817/GRB 170817A \citep{Abbott2017PhRvL.119p1101A}. If the
merger product is a magnetar, it can potentially generate repeating
FRBs. Therefore, monitoring the sites of short gamma-ray bursts is
important to test this hypothesis. Our results suggest that
magnetars powering FRBs may be from core-collapses of massive stars,
which also can produce long gamma-ray bursts. Some searches have
been performed
\citep{Men2019MNRAS.489.3643M,Madison2019arXiv190911682M}. However,
no FRB signal is found. Therefore, future searches with more
sensitive radio telescopes, i.e., FAST \citep{Li18}, are important.
Magnetars from core-collapse of massive stars and BNS mergers will
have distinct host galaxy properties and spatial offset
distributions \citep{Wang19}. Therefore, the localization of a large
sample of FRBs by Australian Square Kilometre Array Pathfinder and
Very Large Array can shed light on the formation channel of
magnetars. The ongoing CHIME and Advanced LIGO/Virgo observations
can refine the present analysis.

\section*{Acknowledgements}
We thank an anonymous referee for constructive and helpful comments.
This work is supported by the National Natural Science Foundation of
China (grant U1831207).



\clearpage

\end{document}